\documentclass[9pt,twocolumn,twoside]{opticajnl}
\journal{opticajournal}

\setboolean{shortarticle}{false}
\doi{}

\title{Off-focus detection for SNR enhancement in coherent single-pixel imaging}

\author[1]{Wei Zhu}
\author[2,*]{Hui Yuan}
\author[1]{Hangbing Guo}
\author[1,3]{Sheng Wang}
\author[1]{Da Tian}

\author[1]{Benwen Chen}
\author[1,3]{Kebin Fan}
\author[1,3]{Caihong Zhang}
\author[1,3]{Huabing Wang}
\author[1,3]{Biaobing Jin}
\author[1,3]{Jian Chen}
\author[1,3]{Peiheng Wu}
\author[2]{Hartmut G. Roskos}
\author[1,3,*]{Jingbo Wu}

\affil[1]{Research Institute of Superconductor Electronics (RISE), School of Electronic Science and Engineering, Nanjing University, Nanjing 210023, China}
\affil[2]{Physikalisches Institut, Goethe-Universität Frankfurt am Main, Frankfurt am Main, Germany}
\affil[3]{Purple Mountain Laboratories, Nanjing 211111, China}

\affil[*]{Corresponding authors: yuan@physik.uni-frankfurt.de, jbwu@nju.edu.cn}

\begin{abstract}
Coherent single-pixel imaging (CSPI) enables amplitude- and phase-resolved computational imaging with a single coherent detector, offering particular advantages in spectral regimes where high-performance detector arrays remain unavailable. In conventional Fourier-plane CSPI, the detector is typically placed at the focal point to maximize the collected power.  However, this focal position does not necessarily maximize the reconstruction fidelity for image reconstruction in the presence of system nonidealities. Here, we introduce detector position as an additional degree of freedom for CSPI optimization. We show that appropriately selected off-focus Fourier-plane positions can improve reconstruction performance by suppressing fluctuation-sensitive coherent backgrounds while preserving spatially encoded signals. Experimentally, using a programmable liquid-crystal terahertz spatial light modulator and coherent heterodyne detection, we demonstrate improved amplitude and phase reconstruction from selected off-focus channels compared with focal detection. A symmetric off-focus pair with complementary transverse-wave-vector weighting further enhances image fidelity, contrast, and compressive-sampling robustness through coherent channel fusion. These results establish detector-position engineering as a new strategy for enhancing the robustness and information efficiency of CSPI.
\end{abstract}

\setboolean{displaycopyright}{false}

\begin{document}

\maketitle

\section{Introduction}
Computational imaging has reshaped image formation by combining physical modulation with computational reconstruction\cite{ref01_mait2018computational,ref02_barbastathis2019deep}. Among various computational imaging approaches, single-pixel imaging (SPI) offers a distinct strategy that encodes spatial information through programmable modulation\cite{ref03_watts2014compressive,ref04_li2022dualcolor} and reconstructs images from sequential measurements with a single detector\cite{ref05_edgar2019principles,ref06_gibson2020review,ref07_duarte2008single}. This architecture avoids the need for pixelated detector arrays and is therefore attractive in spectral regimes where high-performance detector arrays remain difficult to realize, such as the terahertz (THz) band\cite{ref08_guerboukha2018realtime,ref09_shrekenhamer2013terahertz,ref10_stantchev2016noninvasive,ref11_stantchev2020realtime}. Continued advances in modulation techniques\cite{ref12_wu2020liquid,ref13_liu2021programmable,ref14_li2023modulo,ref15_shen2024multibit,ref16_zhu2025highcontrast} and reconstruction algorithms\cite{ref17_zhang2015fourier,ref18_zhang2017hadamard,ref19_sun2017russian,ref20_hahamovich2021megahertz,ref21_bian2018comparison} have established SPI as a versatile platform for spectral\cite{ref22_welsh2013fullcolor}, three-dimensional\cite{ref23_sun2013three}, and phase-sensitive imaging\cite{ref24_ota2018complex,ref25_hou2021complex}.

SPI optimization has mainly focused on the design of spatial modulation patterns, sampling strategies, and reconstruction algorithms, while the detector position is generally regarded as a fixed component determined by the optical configuration. In Fourier-plane SPI, this assumption naturally leads to placing the detector at the focal point, where the collected power is maximized\cite{ref26_goodman2017fourier}. However, in coherent imaging systems, maximizing detected power does not necessarily maximize reconstruction performance. The measured field contains not only the desired modulated signals but also coherent background, leakage components, and fluctuations arising from system instability, all of which can influence the effective signal-to-noise ratio (SNR) and reconstruction fidelity\cite{ref27_liu2025degradations,ref28_wang2023fluctuation,ref29_pastuszczak2021differential,ref30_gross2008noise,ref31_gross2007sensitivity}. Therefore, detector position may affect coherent measurement quality not only by changing the collected power, but also by modifying the contribution of coherent backgrounds and system fluctuations.

This issue has remained largely unexplored because Fourier-plane measurements and CSPI employ different sampling strategies. In conventional Fourier-plane measurements, the Fourier-space field is typically recorded as a continuous distribution, where spatial information is obtained from the full Fourier representation rather than from individual detection channels\cite{ref32_yuan2023sixhundred,ref33_yuan2025status}. In contrast, coherent single-pixel imaging samples selected Fourier channels through sequential measurements, where different detection positions may exhibit distinct sensitivities to useful modulation signals and coherent imperfections\cite{ref34_clemente2013holography,ref35_zhao2019complex}. Whether detector-position selection can therefore be exploited to improve CSPI reconstruction remains an open question.

Here, we establish detector position as an additional degree of freedom for Fourier-plane coherent Hadamard SPI. By treating different detector locations as distinct coherent readout channels, we develop a detector-position-dependent channel model that links reconstruction performance to the balance between useful modulated signals and coherent background contributions under practical system imperfections. The model reveals that off-focus detection can outperform focal readout when the reduction of background-sensitive fluctuations compensates for the loss of collected signal. We validate this framework through both simulations and experiments using a liquid-crystal (LC) THz spatial light modulator (SLM) and coherent heterodyne detection. Furthermore, we extend detector-position optimization to multi-channel fusion by combining off-focus channels with complementary transverse-wave-vector responses, achieving improved image fidelity and robustness under compressive sampling.

\section{Theory and Analysis}
\subsection{Concept of multi-point Fourier-plane CSPI}
\begin{figure*}[ht]
\centering
\includegraphics[width=0.9\linewidth]{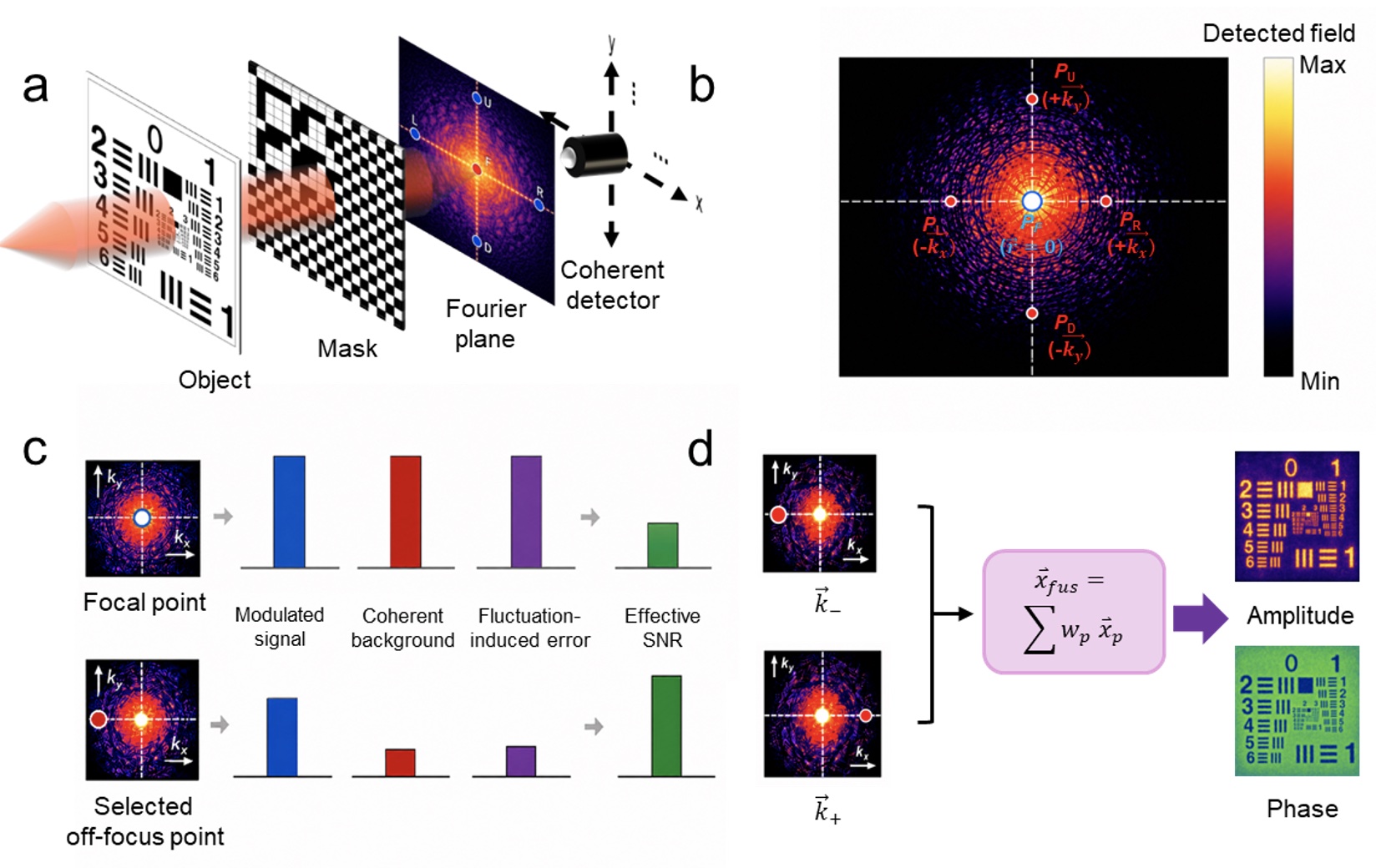}
\caption{Concept of detector-position diversity and symmetric off-focus fusion in Fourier-plane CSPI. (a) Coherent Fourier-plane Hadamard SPI with a movable single-pixel coherent detector. (b) Selection of focal and off-focus detector positions in the Fourier plane, corresponding to distinct transverse-wave-vector channels. (c) Comparison of focal and off-focus readout channels in terms of modulated signal, coherent background, fluctuation-sensitive noise, and effective SNR. (d) Fusion of a symmetric off-focus pair with opposite transverse-wave-vector weightings for amplitude and phase reconstruction.}
\label{fig:1}
\end{figure*}

Fig. \ref{fig:1} summarizes the concept of detector-position diversity in Fourier-plane Hadamard CSPI. Fig. \ref{fig:1}a shows the system configuration, where a Hadamard-coded object field is mapped onto the Fourier plane and measured by a movable coherent single-pixel detector. Fig. \ref{fig:1}b identifies the focal and off-focus detection positions, with each position corresponding to a distinct transverse-wave-vector channel. Fig. \ref{fig:1}c illustrates that the focal point, although providing the strongest detected field, does not necessarily yield the optimal SNR, because coherent background and fluctuation-sensitive noise can reduce the reconstruction-relevant effective SNR. Fig. \ref{fig:1}d further illustrates the fusion of a symmetric off-focus pair with opposite transverse-wave-vector weightings. In the present experiment, PL and PR, form such a complementary pair and are coherently combined for amplitude and phase reconstruction.

\subsection{Detector-position-dependent effective SNR model}

For a detector located at the Fourier-plane position $p$, the coherent Hadamard measurement under the $i$th pattern can be written as
\begin{equation}
y_{i,p} = \sum_{j} H_{ij} x_j h_p(\mathbf{r}_j) + n_{i,p},
\label{eq:1}
\end{equation}
where $H_{ij}$ is the Hadamard modulation coefficient, $x_j$ is the complex object response at the $j$-th pixel, $\mathbf{r}_j$ is its spatial coordinate, $h_p(\mathbf{r}_j)$ is the detector-position-dependent coherent transfer function, and $n_{i,p}$ is measurement noise. Defining the position-weighted object response as $x_{j,\mathbf{p}} = x_j h_{\mathbf{p}}(\mathbf{r}_j)$, ~\eqref{eq:1} retains the standard linear Hadamard form,
\begin{equation}
y_{i,\mathbf{p}} = \sum_{j} H_{ij} x_{j,\mathbf{p}} + n_{i,\mathbf{p}}.
\label{eq:1b}
\end{equation}
Therefore, each detector position with a sufficiently strong and nonvanishing transfer response defines a valid coherent SPI measurement channel, and the corresponding complex image can be reconstructed using the same Hadamard inversion. The reconstructed field is the object response weighted by the position-dependent transfer function. When $h_{\mathbf{p}}(\mathbf{r})$ is known or calibrated and remains sufficiently strong over the field of view, the object response can, in principle, be recovered through calibrated compensation.

At the focal point, $h_{\mathbf{p}}(\mathbf{r})$ provides an approximately uniform coherent phase weighting across the object, whereas an off-focus position introduces a spatially varying phase weighting. Changing the detector position does not introduce additional Hadamard measurements within a single channel. Instead, it modifies the complex spatial weighting of the readout channel. Under ideal and noise-free conditions, both focal and off-focus channels can support image reconstruction after position-dependent calibration. Their practical performance, however, can differ because detector position also changes the useful modulated signal, coherent background, and sensitivity to system fluctuations. The coherent linear SPI model, its comparison with the scalar model, and the detailed multi-position detection model are presented in Supplementary Sections 1 and 2.A.

To determine which detector position is more favorable for image reconstruction, the measured response at position $p$ can be separated into a useful Hadamard-modulated component, a residual coherent background, and additive noise:
\begin{equation}
y_p = s_p + b_p + n_0.
\label{eq:2}
\end{equation}
In ~\eqref{eq:2}, $s_p$ represents the modulated signal, $b_p$ represents the coherent background including common-field contribution, zero-order leakage, and residual unmodulated leakage, and $n_0$ denotes additive detector or electronic noise. The background term $b_p$ is not noise by itself if it is stable and perfectly calibrated. However, in a coherent system, amplitude and phase fluctuations can act multiplicatively on the total coherent field. Including a small relative amplitude fluctuation $\delta a$ and a small phase fluctuation $\delta\phi$, the measured field becomes:
\begin{equation}
y_p = (1 + \delta a)\, e^{i\delta\phi} (s_p + b_p) + n_0.
\label{eq:3}
\end{equation}
For $|\delta a| \ll 1$ and $|\delta\phi| \ll 1$ rad, ~\eqref{eq:3} can be linearized as:
\begin{equation}
y_p \approx s_p + b_p + \delta a (s_p + b_p) + i\delta\phi (s_p + b_p) + n_0.
\label{eq:4}
\end{equation}
~\eqref{eq:4} shows that the fluctuation-induced error scales with the total coherent field. Consequently, even when the static background is reduced by reference subtraction, complementary Hadamard measurements, or calibration, its fluctuation-coupled residual can still degrade reconstruction.

We define the reconstruction-relevant effective SNR at detector position $p$ as
\begin{equation}
\mathrm{SNR}_{eff}(p) = \frac{P_{sig}(p)}{\sigma_0^2 + \sigma_{mult}^2(p)},
\label{eq:5}
\end{equation}
where $P_{sig}(p)$ is the power of the useful Hadamard-modulated signal, $\sigma_0^2$ is the additive noise variance, and $\sigma_{mult}^2(p)$ denotes the effective variance associated with multiplicative amplitude and phase fluctuations. According to the first-order drift model, this multiplicative fluctuation term can be expressed as
\begin{equation}
\sigma_{mult}^2(p) \propto |s_p + b_p|^2 (\sigma_a^2 + \sigma_\phi^2) \approx \eta_p |b_p|^2 (\sigma_a^2 + \sigma_\phi^2).
\label{eq:6}
\end{equation}
where $\sigma_a$ and $\sigma_\phi$ are the standard deviations of the relative amplitude fluctuation and phase fluctuation, and $\eta_p$ is a residual-coupling factor that accounts for the fraction of background-related fluctuations remaining after differential measurement and calibration.

This model explains why the focal point is not necessarily optimal. Although the focal channel usually provides a larger useful modulated response, it may also contain a stronger coherent background and hence greater fluctuation-induced error. Conversely, an off-focus channel can provide a higher effective SNR when
\begin{equation}
\frac{P_{sig}(off)}{\sigma_0^2 + \sigma_{mult}^2(off)} > \frac{P_{sig}(focal)}{\sigma_0^2 + \sigma_{mult}^2(focal)}.
\label{eq:7}
\end{equation}
~\eqref{eq:7} defines the condition under which off-focus detection outperforms focal readout: the reduction in fluctuation-sensitive coherent contributions must outweigh the loss of useful modulated signal. Thus, the optimal Fourier-plane detection position is determined by effective SNR rather than by the maximum collected field amplitude. The effects of slow amplitude and phase fluctuation and finite on-off ratio of the SLM on the effective SNR is analyzed in Supplementary Section 2.B.

\subsection{Experimental-informed noise modeling}
To connect the effective-SNR model with the experiment, we performed repeated reference measurements at the focal and selected off-focus positions in the Fourier-plane. These measurements provide channel-dependent estimates of the reference-field amplitude and its amplitude and phase fluctuations, which serve as experimental proxies for the background-related terms in the effective-SNR model. The relative amplitude fluctuation was calculated from the standard deviation of the repeated reference amplitudes normalized by their mean, while the phase fluctuation was obtained from the standard deviation of the unwrapped repeated reference phases. These experimentally estimated quantities correspond to the background and fluctuation terms as shown in ~\eqref{eq:4} and were used to construct the realistic-noise simulation. The detailed construction of the reference-derived realistic-noise simulation is described in Supplementary Section 2.C.

Fig. \ref{fig:2} summarizes the experiment-informed realistic-noise simulation based on these estimated channel parameters. Fig. \ref{fig:2}a compares the reference-field amplitude and the relative amplitude and phase fluctuations at the focal and off-focus positions. These parameters were incorporated into the coherent Hadamard forward model to so that different detection channels could be evaluated using the same coding basis and reconstruction algorithm. The reference measurements show that the focal position contains the largest coherent background, whereas the off-focus positions exhibit reduced background levels together with different amplitude and phase fluctuation. According to ~\eqref{eq:5}, a larger detected field does not necessarily lead to a higher reconstruction SNR when it is accompanied by stronger multiplicative fluctuations.
\begin{figure*}[ht]
\centering
\includegraphics[width=0.8\linewidth]{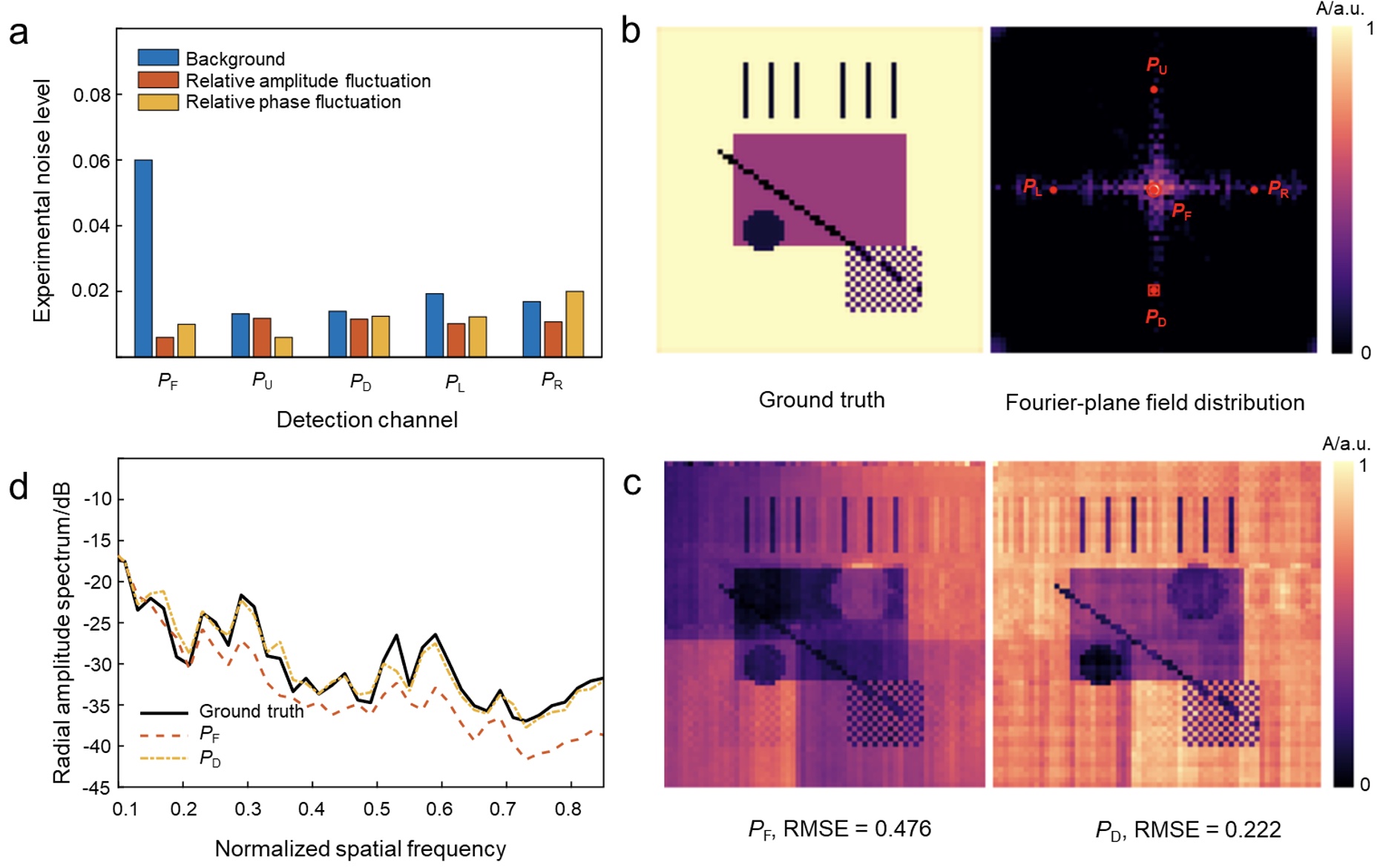}
\caption{Realistic-noise simulation of focal and off-focus coherent detection channels. (a) Experimentally estimated channel non-idealities, including coherent background, relative amplitude fluctuation, and phase fluctuation. (b) Ground-truth amplitude and Fourier-plane field distribution of the simulation object. (c) Reconstructed amplitudes from $P_F$ and $P_D$. (d) Radial averaged spatial spectra of the reconstructed amplitude images, showing the preservation of mid- and high-spatial-frequency components.}
\label{fig:2}
\end{figure*}

Fig. \ref{fig:2}b illustrates the ground-truth amplitude and Fourier-plane field distribution of the simulation object. Then Fig. \ref{fig:2}d gives the reconstructed amplitude of the focal channel and the selected off-focus channel, respectively. It can be concluded that the focal channel reconstruction is more strongly affected by the measured background and fluctuation levels, resulting in blurred edges and reduced fine features. In contrast, the selected off-focus channel retains sufficient modulated signal while reducing the fluctuation-sensitive coherent background, resulting in a sharper reconstruction. The amplitude RMSE is defined as
\begin{equation}
RMSE = \sqrt{\frac{1}{N}\sum_{j=1}^{N}\left(A_{rec,j}-A_{true,j}\right)^{2}},
\label{eq:8}
\end{equation}
where $A_{rec}$ and $A_{true}$ denote the normalized reconstructed and ground-truth amplitude distributions, respectively. The amplitude RMSE decreases from 0.476 for the focal channel to 0.222 for the selected off-focus channel. The radial averaged spatial spectra in Fig. \ref{fig:2}d further shows that the off-focus reconstruction preserves mid- and high-spatial-frequency components, with the high-frequency amplitude error reduced from 0.594 to 0.239. These results are consistent with the effective-SNR model, indicating that reduced background-coupled fluctuations can compensate for the lower detected field at the selected off-focus position.

\subsection{Channel selection and fusion principle}

The single-channel analysis shows that detector-position optimization can improve the effective SNR, but each off-focus channel also carries a position-dependent complex weighting and channel-specific reconstruction error. In particular, symmetrically located off-focus positions provide opposite transverse-wave-vector weightings, suggesting that they may provide complementary coherent readout responses. This motivates the fusion of selected detector-position-dependent channels.

After detector-position-dependent calibration, the reconstruction from the $p$-th channel can be written as $\hat{x}_p = x + e_p$, where $x$ is the common object response and $e_p$ is the channel-dependent reconstruction error, a weighted fusion over a selected channel set $\Omega$ gives
\begin{equation}
\hat{x}_{fus} = \sum_{p \in \Omega} w_p \hat{x}_p = x + \sum_{p \in \Omega} w_p e_p, \qquad \sum_{p \in \Omega} w_p = 1.
\label{eq:9}
\end{equation}
Thus, fusion preserves the common object term, while the residual error is determined by the weighted combination of channel-dependent errors. The corresponding fusion-error variance is
\begin{equation}
\sigma_{fus}^{2} = \mathbf{w}^{H} \mathbf{C}_{\Omega} \mathbf{w}, \qquad C_{pq} = \langle e_p e_q^{*} \rangle, \quad p,q \in \Omega.
\label{eq:10}
\end{equation}
where $\mathbf{C}_{\Omega}$ is the inter-channel error covariance matrix. ~\eqref{eq:10} shows that the benefit of fusion is determined by both the individual channel errors and the correlations among them. Adding a weak or highly correlated channel may provide little benefit or even degrade the reconstruction.

In Fourier-plane coherent detection, useful fusion requires both sufficient modulated signal and complementary phase weighting. A symmetric off-focus pair can provide opposite transverse-wave-vector weightings while retaining measurable Hadamard responses. These considerations suggest that signal strength and transverse-wave-vector diversity can serve as practical criteria for channel selection.

\section{Experiment and Discussion}
\subsection{Coherent Fourier system and pre-scan}

\begin{figure}[ht]
\centering
\includegraphics[width=\linewidth]{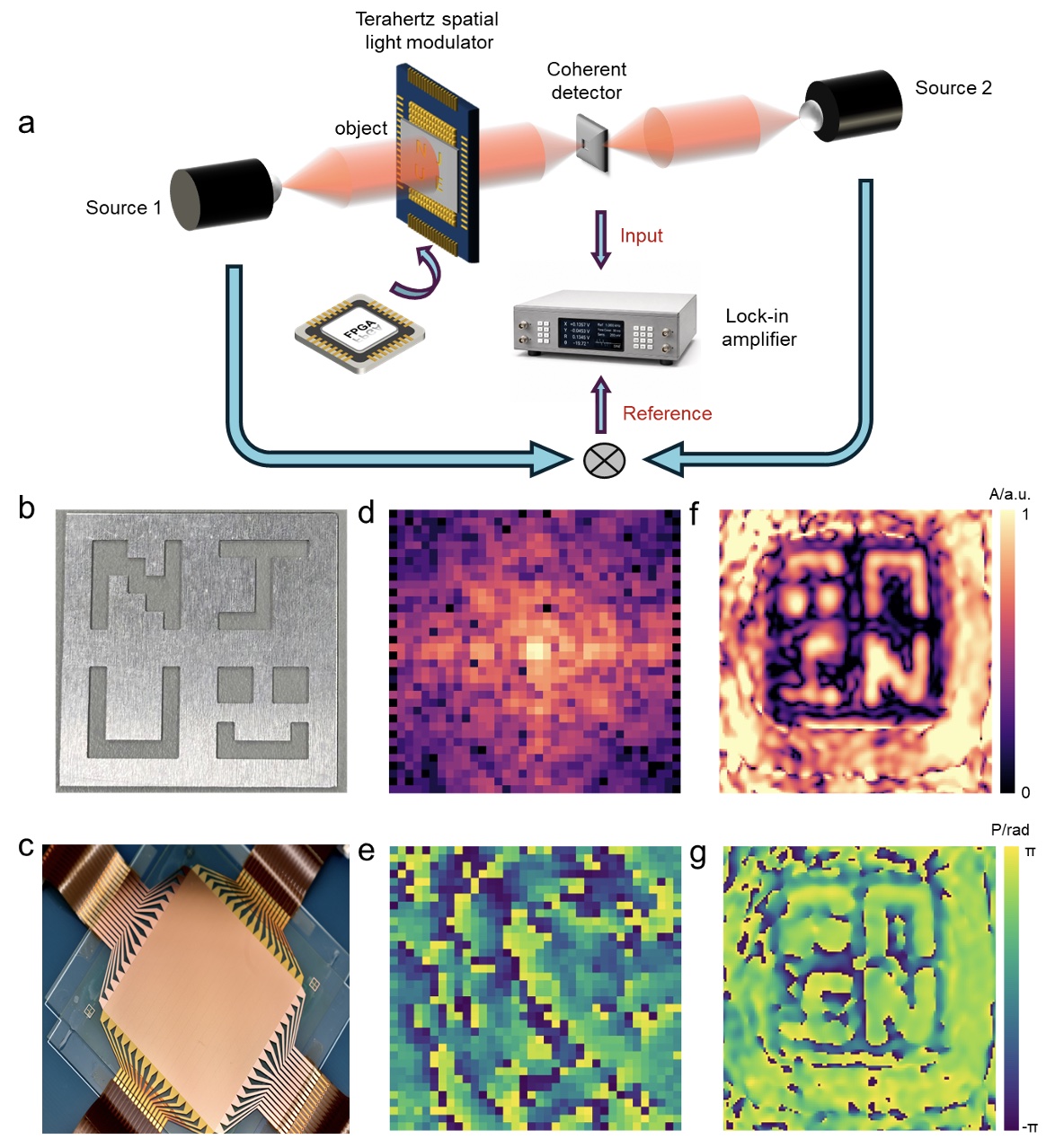}
\caption{Coherent THz imaging system and Fourier-plane scanning reconstruction. (a) Schematic of the coherent heterodyne THz imaging system. (b) Photograph of the metallic test object. (c) Photograph of the LC-SLM. Measured Fourier-plane (d) amplitude and (e) phase distributions obtained by raster-scanning using the coherent detector. Reconstructed object-plane (f) amplitude and (g) phase distributions obtained from the measured complex Fourier-plane field.}
\label{fig:3}
\end{figure}

To experimentally validate the detector-position-dependent effective-SNR model and the channel-selection principle described above, we implemented a coherent Fourier-plane Hadamard SPI system with a movable single-pixel detector, as shown in Fig.~\ref{fig:3}a. Details are provided in Methods.

The LC SLM operates in a transmission mode and was positioned immediately behind the metallic object, forming a combined object-modulation plane. Driven by an FPGA, it displayed complementary Hadamard amplitude patterns for CSPI. With all pixels set to the high-transmission state, the LC-SLM enabled coherent Fourier-plane raster scanning, while pixel-wise electrical addressing provided programmable amplitude modulation. The in-house fabricated device showed a modulation depth exceeding 85\% near 310 GHz. The detailed fabrication process, unit-cell design, and spectral modulation performance of the LC-SLM are provided in Supplementary Section 3.

Before performing CSPI measurements, we conducted a raster scan of the Fourier plane with all pixels of SLM set to the high-transmission state and without Hadamard modulation. This measurement was used to characterize the complex Fourier-plane response of the object-SLM-detector system and to identify suitable regions for subsequent detector-position selection. Figs. \ref{fig:3}b and \ref{fig:3}c show the metallic test object and the LC-SLM, respectively. The metallic features of the test object have a linewidth of approximately 2 mm, close to the experimentally demonstrated lateral resolution of a related 310-GHz Fourier-imaging system\cite{ref36_yuan2019threedimensional}. The coherent detector was raster-scanned across the Fourier plane to record the complex field distribution. The measured Fourier-plane amplitude and phase maps are shown in Figs. \ref{fig:3}d and \ref{fig:3}e, respectively. The amplitude map exhibits a horizontally extended field distribution, indicating that the left and right off-focus regions retain appreciable field strength for subsequent Hadamard readout.

The measured complex Fourier-plane field was then numerically transformed back to the object plane, yielding the amplitude and phase reconstructions shown in Figs. \ref{fig:3}f and \ref{fig:3}g. The reconstructed amplitude image confirms that the LC-SLM in the uniform high-transmission state preserves sufficient THz throughput for object recovery, while the phase image provides an additional assessment of the coherent phase response. Together, these measurements characterize the complex-field response of the system and identify suitable focal and off-focus regions for the subsequent Hadamard measurements.

\subsection{Detector-position-dependent coherent SPI}

Based on the measured Fourier-plane complex-field distribution, five detection positions were selected for the subsequent Hadamard single-pixel imaging measurements: the focal position and four off-focus positions along the horizontal and vertical directions. The off-focus displacement was chosen to provide sufficient spatial-frequency separation from the focal channel while remaining within the region of reliable coherent response. Further details of the selected positions, detected-field distribution, and spatial-frequency separation are provided in Supplementary Section 2.D.
During the Hadamard measurements, the SLM was operated with complementary Hadamard amplitude patterns, while the detector was sequentially positioned at the five selected Fourier-plane locations, denoted $P_{F}$, $P_{U}$, $P_{D}$, $P_{L}$, and $P_{R}$, corresponding to the focal, upper, lower, left, and right positions, respectively.

For each detection position, the amplitude and phase measured under the positive and negative Hadamard masks were combined to form complex-valued differential measurements. The same Hadamard coding basis and reconstructed using the same total-variation (TV)-regularized algorithm\cite{ref37_rudin1992total,ref38_wang2008alternating} for all five channels. Therefore, the differences among the reconstructed images arise primarily from the detection-position-dependent coherent channel response rather than changes in the coding basis or reconstruction algorithm.

The structural similarity index measure (SSIM)\cite{ref39_wang2004ssim} and contrast-to-noise ratio (CNR)\cite{ref40_timischl2015cnr} were used to evaluate the structural fidelity and object--background contrast of the amplitude and phase reconstructions, respectively. Details are provided in Methods.

\begin{figure*}[ht]
\centering
\includegraphics[width=0.8\linewidth]{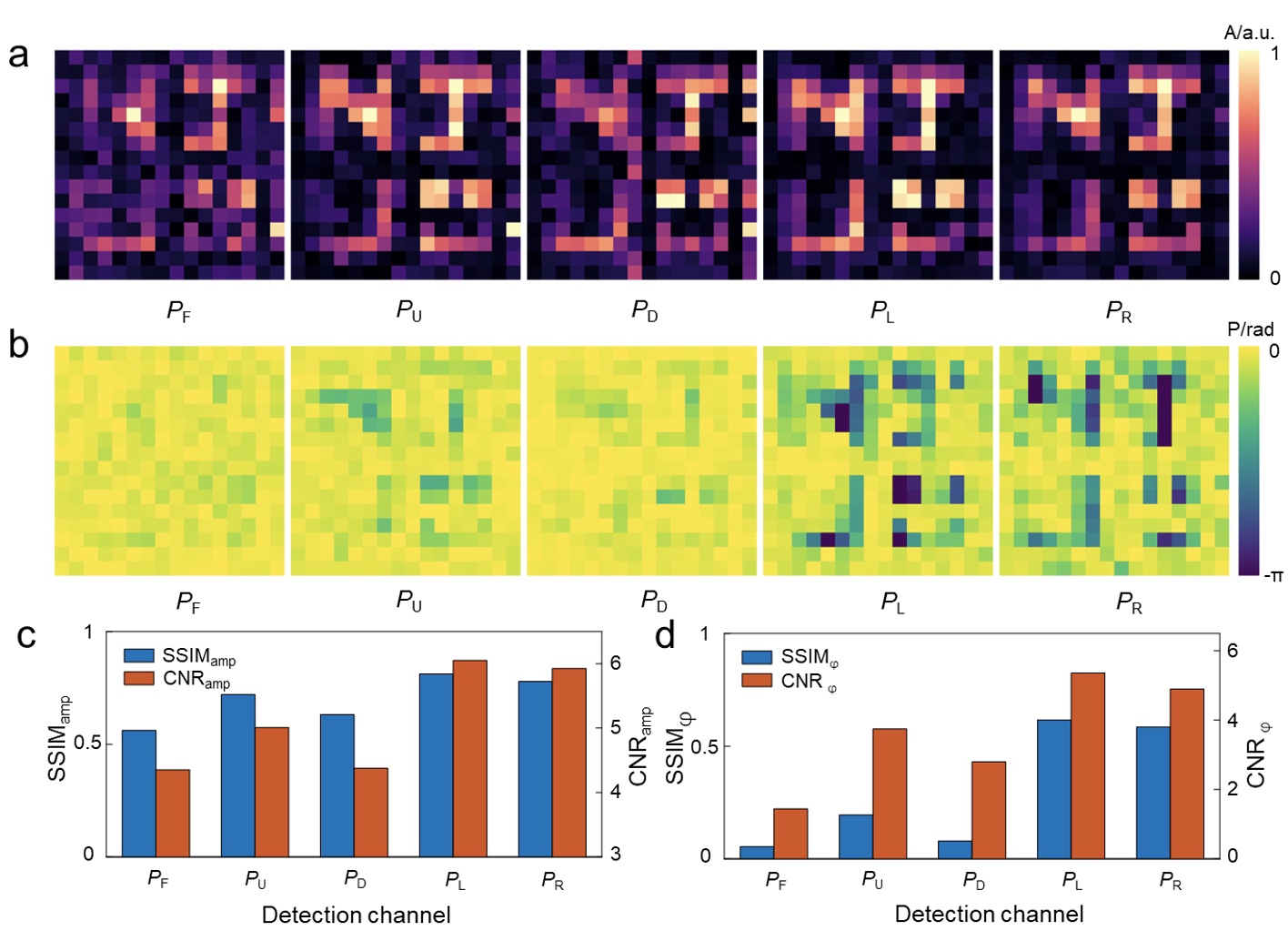}
\caption{Detector-position-dependent CSPI. (a) Amplitude and (b) phase reconstructions from five Fourier-plane detector positions: $P_F$ (Focal), $P_U$ (Up), $P_D$ (Down), $P_L$ (Left), and $P_R$ (Right). SSIM and CNR evaluations of the (c) amplitude and (d) phase reconstructions. }
\label{fig:4}
\end{figure*}

Figs. \ref{fig:4}a and \ref{fig:4}b compare the amplitude and phase reconstructions obtained at the five detection positions, and Figs. \ref{fig:4}c and \ref{fig:4}d summarize the corresponding SSIM and CNR values. The focal channel yields a recognizable but relatively weak reconstruction, characterized by reduced object contrast and pronounced background fluctuation. By contrast, the off-focus channels provide clearer recovery of the target structure. In the amplitude reconstructions, the $P_{U}$, $P_{L}$, and $P_{R}$ channels show improved object visibility compared with the focal channel, while the $P_{L}$ channel gives the highest overall SSIM and CNR. A similar trend is observed in the phase reconstructions, where the focal channel exhibits weaker phase-related object features, whereas the off-focus channels, especially at $P_{L}$, and $P_{R}$, provide substantially clearer phase recovery.

These results agree with the effective-SNR model developed above, as selected off-focus positions achieve improved reconstruction quality despite collecting a weaker absolute field than the focal position.

\subsection{Multi-channel fusion of selected off-focus coherent reconstructions}

To further improve the reconstruction beyond a single detection position, we investigated the fusion of selected off-focus coherent readout channels. We first consider two-channel fusion, which is the simplest configuration for testing detector-position complementarity while limiting the additional acquisition time required by sequential Hadamard measurements. To benchmark the selected pair against other two-channel combinations, we jointly considered individual-channel reconstruction quality and inter-channel error redundancy.

For each reconstructed quantity $m$, where $m = A$ and $m = \phi$ denote amplitude and phase, respectively, the quality score of channel $p$ is defined as:
\begin{equation}
Q_{p}^{m} = \alpha \widetilde{SSIM}_{p}^{m} + (1-\alpha)\widetilde{CNR}_{p}^{m}.
\label{eq:15}
\end{equation}
where the normalized SSIM and CNR quantify structural fidelity and object--background contrast. The overall channel-quality score is
\begin{equation}
Q_{p} = \frac{Q_{p}^{A} + Q_{p}^{\phi}}{2}.
\label{eq:16}
\end{equation}
The redundancy between channels $p$ and $q$ is quantified by the mean correlation of their amplitude- and phase-error maps:
\begin{equation}
\rho_{pq} = \frac{\rho_{pq}^{A} + \rho_{pq}^{\phi}}{2},
\label{eq:17}
\end{equation}
where $\rho_{pq}^{A}$ and $\rho_{pq}^{\phi}$ are the Pearson correlation coefficients of the amplitude- and phase-error maps, respectively. The benchmark score for a candidate pair is then defined as
\begin{equation}
B_{pq} = \frac{Q_p + Q_q}{2}\left(1 - \rho_{pq}\right).
\label{eq:18}
\end{equation}
which favors pairs with both high reconstruction quality and low error redundancy.

\begin{figure}[ht]
\centering
\includegraphics[width=\linewidth]{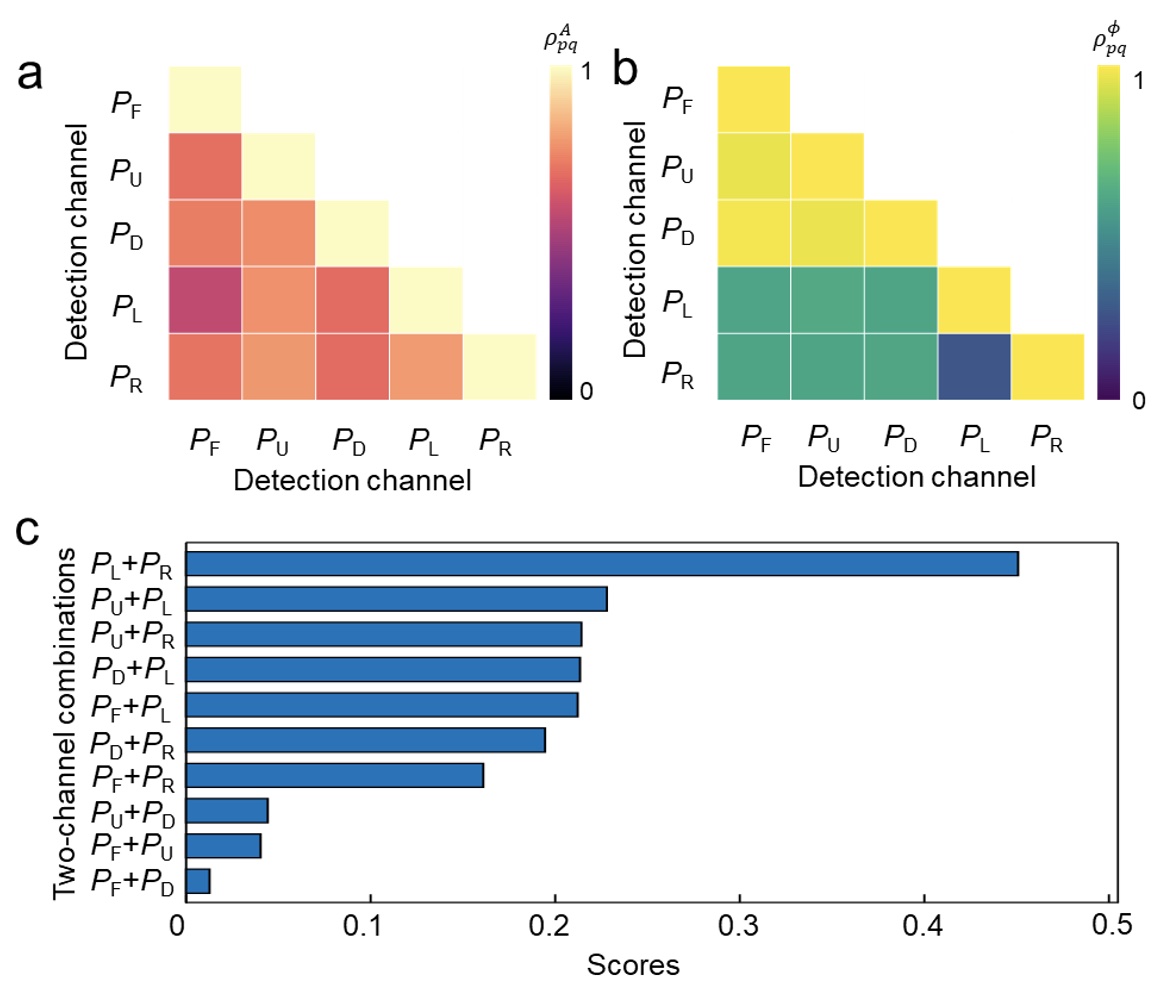}
\caption{Quantitative comparison of candidate two-channel fusion pairs. (a) Correlation matrix of the amplitude reconstruction-error maps. (b) Correlation matrix of the phase reconstruction-error maps. (c) Benchmark scores for the candidate channel pairs.}
\label{fig:5}
\end{figure}

Fig. \ref{fig:5} compares the candidate channel pairs. Although the $P_L$--$P_R$ pair does not exhibit the lowest amplitude-error correlation, it gives the lowest phase-error correlation of 0.2715 and the lowest combined error correlation of 0.5266. After accounting for individual-channel quality, the $P_L$--$P_R$ pair achieves the highest benchmark score of 0.4508, compared with 0.2280 for the next-best $P_U$--$P_L$ pair. Under this benchmark, the symmetric $P_L$--$P_R$ pair therefore provides the best balance between reconstruction quality and error complementarity.

\begin{figure*}[ht]
\centering
\includegraphics[width=\linewidth]{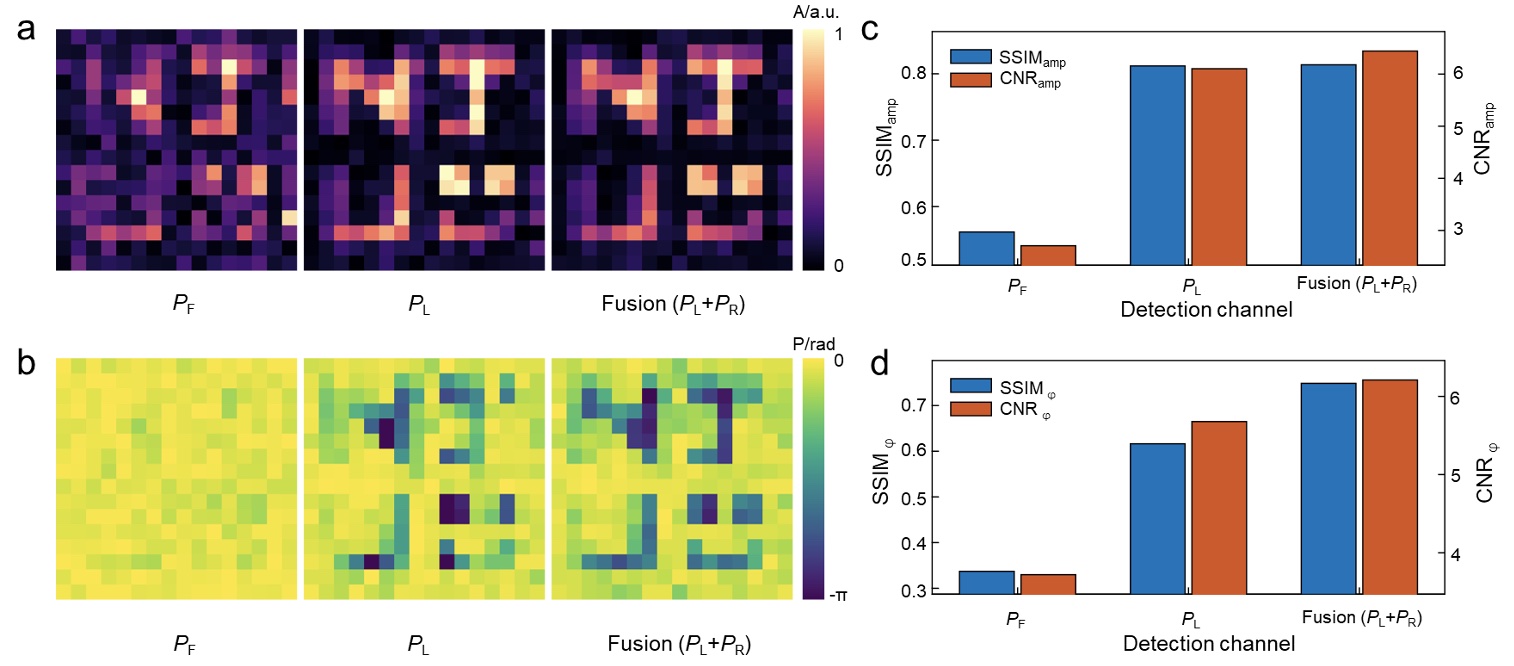}
\caption{Fusion of the symmetric off-focus coherent readout pair. (a) Amplitude and (b) phase reconstructions obtained from the focal channel $P_F$, the representative $P_L$ channel, and the fused $P_L – P_R$. (c) Amplitude SSIM and CNR. (d) Phase SSIM and CNR.}
\label{fig:6}
\end{figure*}

Based on the physically selected $P_L$--$P_R$ pair and the benchmark analysis above, the two off-focus reconstructions were fused with equal weights using the model in ~\eqref{eq:9}. Fig. \ref{fig:6} compares the reconstructions obtained from the focal channel $P_F$, the representative off-focus channel $P_L$, and the fused $P_L$--$P_R$ pair. For amplitude, the $P_F$ channel gives a recognizable but low-contrast image, with an amplitude SSIM of 0.5617 and CNR of 2.7067. The $P_L$ channel substantially improves these values to 0.8118 and 6.1026, respectively. Fusion further increases the CNR to 6.4419 while maintaining a similarly high SSIM of 0.8135. Thus, the amplitude improvement arises primarily from selecting a high-quality off-focus channel, with fusion providing an additional enhancement in object--background contrast.

The fusion benefit is more pronounced for phase reconstruction. While $P_L$ already provides a substantial improvement over $P_F$, fusing $P_L$ and $P_R$ further increases the phase SSIM and CNR by approximately 21\% and 20\%, respectively. This enhancement is consistent with the low phase-error correlation identified in Fig. \ref{fig:5}, indicating effective suppression of complementary phase artifacts through symmetric off-focus fusion.

\section{Discussion}
These results address a question that has received limited attention in coherent Fourier-plane SPI: whether the focal point is necessarily the optimal detection position for image reconstruction. By treating detector position as an optimizable degree of freedom, we show that selected off-focus channels can outperform the focal channel despite collecting a weaker absolute field, and that symmetric off-focus fusion can further improve reconstruction through channel complementarity. This advantage is maintained under reduced Hadamard sampling, as shown in Supplementary Fig. S4,
indicating that detector-position engineering is relevant not only to full-sampling reconstruction but also to compressive acquisition. The present work therefore opens a broader design space beyond the conventional focal-point readout configuration.
The present study nevertheless remains an initial demonstration. The dependence of the optimal detection position on object type, sample complexity, and noise conditions has not yet been systematically investigated, and a more detailed analysis of coherent background, phase fluctuations, and other noise sources is still required. In addition, the current experiments are performed at a single operating frequency. Extending the method to broadband measurements and other spectral regions will be important for evaluating its generality and practical applicability.

\section{Conclusion}
In conclusion, we establish detector position as an optimizable readout degree of freedom in Fourier-plane CSPI. Although the focal point provides the strongest detected field, selected off-focus positions can yield superior amplitude and phase reconstructions by improving the balance between Hadamard-modulated signal and fluctuation-sensitive background. Moreover, fusion of a symmetric off-focus channel pair further improves reconstruction by exploiting channel complementarity, with particularly clear benefits for phase recovery. These results identify detector-position engineering as a practical and previously underexplored route for improving coherent computational imaging without modifying the coding basis or detection architecture.
\\
\\

\noindent \textbf{APPENDIX A: MATERIALS AND METHODS}

\noindent \textbf{Experimental setup:} A multiplier-chain source was used as the object-illumination beam, which was collimated and transmitted through the metallic test object and the THz LC-SLM. The object was placed in close contact with the SLM surface. The transmitted THz field was then focused onto the Fourier plane, where a coherent detector recorded the local field. A second synchronized multiplier-chain source served as the local oscillator and was coupled to the same detector. In the 310-GHz implementation, the illumination source was driven by a 17.22-GHz synthesizer and multiplied by a factor of 18, while the local-oscillator source was an RPG-Radiometer Physics GmbH multiplier-chain source driven by a slightly frequency-shifted synthesizer. The signal and local-oscillator fields were mixed at an antenna-coupled TeraFET detector, and the difference-frequency signal was demodulated by a PerkinElmer lock-in amplifier with a 50 ms integration time and approximately 60 dB dynamic range using the electrical reference generated from the mixed drive signals of the two synthesizers. Therefore, this heterodyne configuration enables measurement of the complex Fourier-plane field, including both amplitude and phase.

\noindent \textbf{Evaluation indications of reconstructed images:} To evaluate the reconstructed amplitude and phase maps, we used the structural similarity index measure (SSIM) and contrast-to-noise ratio (CNR). SSIM measures the structural agreement between a reconstructed image $I$ and a ground-truth image $G$,
\begin{equation}
SSIM(I,G) = \frac{(2\mu_I\mu_G + C_1)(2\sigma_{IG} + C_2)}{(\mu_I^{2} + \mu_G^{2} + C_1)(\sigma_I^{2} + \sigma_G^{2} + C_2)},
\label{eq:19}
\end{equation}
where $\mu_I$ and $\mu_G$ are the mean values, $\sigma_I^{2}$ and $\sigma_G^{2}$ are the variances, $\sigma_{IG}$ is the covariance, and $C_1$ and $C_2$ are small constants for numerical stability. A higher SSIM indicates better preservation of the object structure. CNR was calculated between the object and background regions as
\begin{equation}
CNR = \frac{|\mu_{obj} - \mu_{bg}|}{\sigma_{bg} + \epsilon},
\label{eq:20}
\end{equation}
where $\mu_{obj}$ and $\mu_{bg}$ are the mean values in the object and background regions, respectively, and $\sigma_{bg}$ is the standard deviation of the background region. A higher CNR indicates stronger object-background contrast and lower background fluctuation.

\begin{backmatter}
\bmsection{Funding} National Natural Science Foundation of China (62331015, 62288101, 62522113, 62275118, 62401252, 62531009, and 92463308), Jiangsu Science Fund for Distinguished Young Scholars (BK20240062), Natural Science Foundation of Jiangsu Province (BK20241225), Key project of Basic Research Program of Jiangsu Province (BK20253015), the Fundamental Research Funds for the Central Universities, the Research Fund for Jiangsu Key Laboratory of Advanced Techniques for Manipulating Electromagnetic Waves, Deutsche Forschungsgemeinschaft (RO 770/48-1), and the Hessian Ministry of Higher Education, Research, Science and the Arts (LOEWE programme, 114110511).

\bmsection{Acknowledgment} W. Z., J. W., H. Y. and H. G. R. conceived the idea. W. Z. designed and fabricated the device. H. G. gave important guidances on hardware control. W. Z. and H. Y. built the experimental setup and performed the measurement. W. Z., J. W., and H. Y. performed the theoretical analysis. S. W., D. T., B. C., K. F., C. Z., H. W., B. J., J. C. and P. W. contributed to constructive discussions on the idea and the paper. W. Z. and J. W. wrote the paper with input from all authors. W. Z, J. W., H. Y., and H. G. R. revised the paper. H. G. R. gave the experimental support. J. W., H. Y., H. G. R. and B. J. supervised the project.

\bmsection{Disclosures} The authors declare no conflicts of interest.

\bmsection{Data Availability Statement} The data that support the finding of this study are available from the corresponding author upon request.

\bmsection{Supplemental document}
See Supplement 1 (provided as an ancillary file with this preprint) for supporting content.
\end{backmatter}

\end{document}